\documentclass[prl,aps,twocolumn,showpacs]{revtex4}
\usepackage{graphicx}
\usepackage{dcolumn}
\usepackage{bm}
\usepackage{amsmath}%
\usepackage{amsfonts}%
\usepackage{amssymb}

\begin{document}

\title{Phase controlled superconducting proximity effect probed by tunneling
spectroscopy}
\author{H. le Sueur, P.\ Joyez, H.\ Pothier, C.\ Urbina, and D.\ Esteve}
\affiliation{Quantronics group, Service de Physique de l'Etat Condens\'{e} (CNRS URA
2464), CEA-Saclay, 91191 Gif-sur-Yvette, France}
\date{\today}

\begin{abstract}
Using a dual-mode STM-AFM microscope operating below $50~\mathrm{mK}$ we
measured the Local Density of States (LDoS) along small normal wires
connected at both ends to superconductors with different phases. We observe
that a uniform minigap can develop in the whole normal wire and in the
superconductors near the interfaces. The minigap depends periodically on the
phase difference. The quasiclassical theory of superconductivity applied to
a simplified 1D model geometry accounts well for the data.
\end{abstract}

\maketitle

The proximity effect at the interface between a normal metal (N) and a superconductor (S) consists
in the weakening of electron pair correlations on the S side and in their appearance on the N side,
which can then acquire superconductor-like properties (for a review, see \cite{Pannetier}). In the
case of diffusive metals, the quasiclassical theory of superconductivity \cite{Belzig} provides a
powerful framework to describe this proximity effect. It predicts that if the N region is smaller
than the electron phase coherence length, it then behaves as a genuine superconductor, \emph{i.e.}
it can carry a supercurrent and there is an energy range $[-\delta,+\delta]$ around the Fermi
energy in which there are no available states for quasiparticles. As $\delta $  is smaller than the
gap $\Delta $ of the bulk superconductor, it has been dubbed the ``minigap'' \cite{Belzig}.  A
remarkable feature is that this minigap is the same everywhere in the structure, \textit{i.e.} both
in the N and in the S electrodes. In particular, this implies a non-zero Local Density of States
(LDoS) in the superconductor between $\delta $ and $\Delta $. This additional density of states,
which can be seen as the evanescent tail of normal electrons undergoing Andreev
reflection~\cite{Pannetier} at the NS interface, dies exponentially when moving into the
superconductor. These predictions are illustrated in Fig.~1a for the case of a one dimensional SNS
structure in which $\delta$ is furthermore predicted to depend periodically on the phase difference
$\varphi $ between the order parameters of the two superconductors \cite{Zhou} and vanishes at
$\varphi =\pi \operatorname{mod}2\pi $, as shown in Fig.~1b. Although spectral properties of
proximity structures
have been partially probed in spatial measurements of the LDoS \cite%
{Gueron,Vinet, Moussy,Gupta,Escoffier} and in transport measurements (refs. in \cite{Pannetier} and
\cite{Dubos,Petrashov,Crosser,Angers}), neither the existence of a uniform minigap nor its phase
modulation have been reported. In this Letter we present high resolution measurements of the phase
and space dependence of the LDoS on both sides of the interfaces of SNS structures, and compare
them to the theoretical predictions.

\begin{figure}[tbph]
\begin{center}
\includegraphics[angle=-90,width=.85\columnwidth,clip]{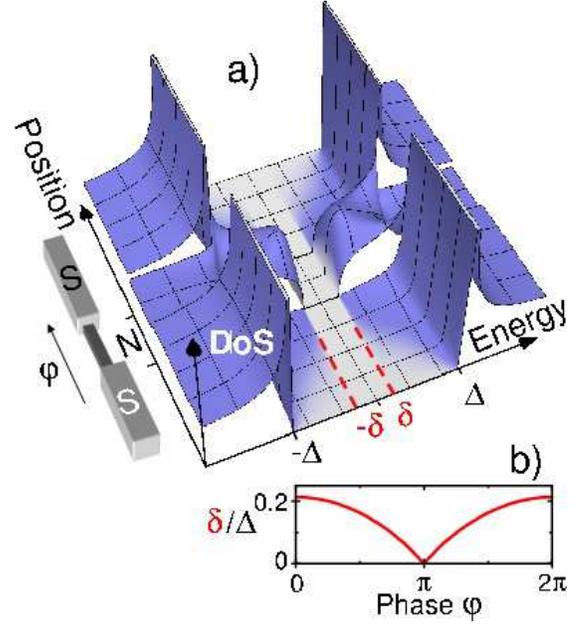}
\end{center}
\caption{(color online) (a) LDoS in a 1-D SNS structure, obtained by solving the Usadel equations
of the quasiclassical theory of superconductivity in diffusive metals, taking into account
scattering at interfaces and geometrical parameters~\protect\cite{AdditionalMaterial}. $\Delta$ is
the gap of the superconductor. A uniform minigap~$\protect\delta$, defining a range of energy
$2\protect\delta$ in which the LDOS is exactly zero, appears across the normal region and in the S
electrodes. In the latter, the finite conductance between $\protect\delta$ and $\Delta$ decreases
exponentially away from the interfaces, and the BCS DoS is eventually recovered. The color
scale was chosen to enhance the visibility of the minigap. (b) Predicted $2%
\protect\pi $-periodic phase modulation of $\protect\delta$.}
\end{figure}

\begin{figure*}[tbph]
\begin{center}
\includegraphics[angle=0,width=2\columnwidth,clip]{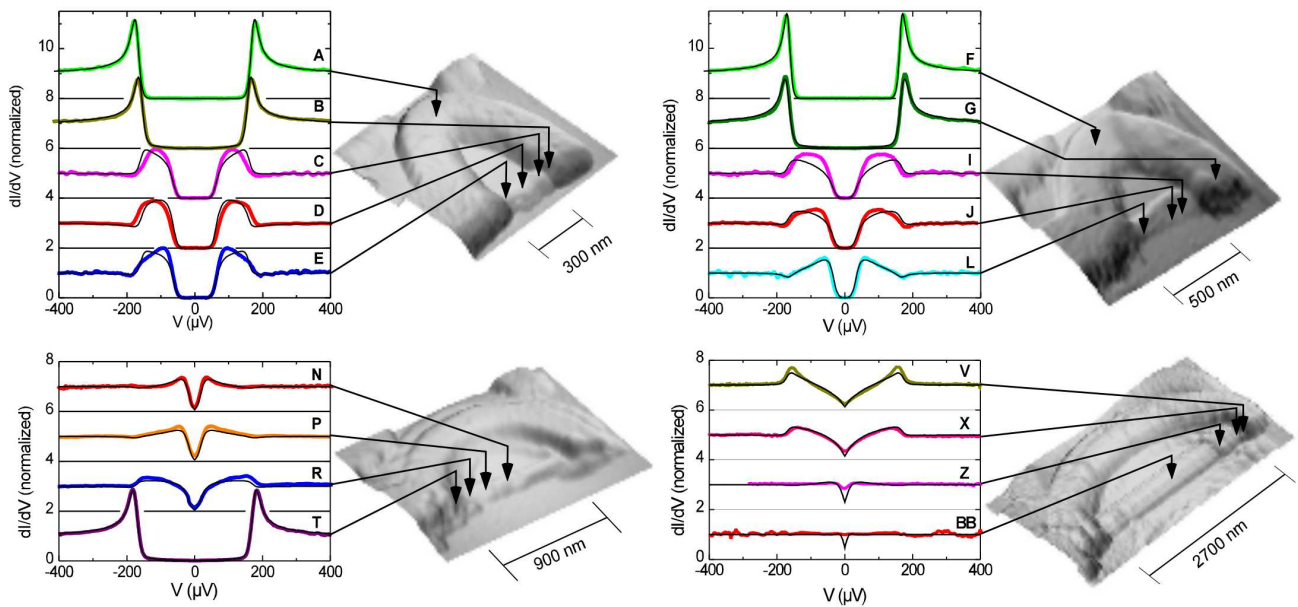}
\end{center}
\caption{(color online) Measured (thick grayed lines) differential conductance taken for zero flux
through the loop at various positions (indicated by the arrows and labeled by capital letters) of
the four SNS structures shown by the AFM images (taken at 35~mK). The length of the silver wire is
respectively (a) 300~nm, (b) 500~nm, (c) 900~nm, (d) 2700~nm. The curves are normalized to 1 at
large voltage and are shifted by integer numbers for clarity. The thin black solid lines correspond
to the model based on the quasiclassical theory of superconductivity introduced in the text and
described more in depth in Ref.~\protect\cite{AdditionalMaterial}.} \label{AFM}
\end{figure*}

In a Scanning Tunneling Microscope (STM), the dependence of the tunneling current $I$ on the
tip-to-sample voltage $V$ directly probes the electronic states available at energy $eV$ in the
electrode just underneath the sharp tip. Variations of the LDoS over small distances can therefore
be accessed taking advantage of the high spatial resolution of the STM. In a measurement with a
normal metal tip, the differential conductance $dI/dV(V)$ at small energies is simply proportional
to the LDoS of the electrode, broadened by an instrumental function which is ideally the derivative
of the Fermi distribution function of the electrons in the tip at temperature $T_{e}$
\cite{Fischer}. This sets
a limit to the energy resolution of such tunneling spectroscopy at about $%
3.5~k_{B}T_{e}$, which can be nevertheless small enough to resolve the
features due to the proximity effect, provided the electrons in the tip are
cooled well below the temperature $\delta /k_{B}$.

In order to perform such measurements on nanocircuits, which contain large insulating areas on
which STMs cannot be operated, we have designed and
built a cryogenic dual mode STM-AFM (Atomic Force Microscope)\thinspace \cite%
{TheseHelene,AdditionalMaterial}, operating with a single metallic tip. This local probe sensor
consists of an electrochemically sharpened tungsten wire \cite{Kulawik} glued on one prong of a
miniature piezoelectric quartz tuning fork. The latter is a high quality
factor electromechanical resonator, which here serves as the AFM transducer %
\cite{Karrai}. Other dual mode instruments are being developed and used
elsewhere~\cite{Senzier,Smit}. The AFM mode is used to acquire detailed topographic images of the
samples which later on allow for accurate positioning of the tip for STM spectroscopy. The
microscope is
mounted in a table-top dilution refrigerator with a base temperature of $%
\sim $35~mK. By itself this is not sufficient to ensure a low electronic temperature $T_{e}$ and
therefore a high energy resolution: it is of critical importance to also shield and filter all the
electrical lines needed to operate the microscope, and to measure the tunnel current with a low
back-action amplifier. For these purposes, we developed microfabricated cryogenic filters
\cite{filters} and a custom low-noise, differential-sensing current amplifier \cite{amp}. After
cooling down, the tip is moved on the sample substrate towards the structures of interest, using a
3-axes capacitance-indexed coarse-positioning system based on stick-slip motion. During this stage,
the microscope is operated in AFM mode, and recognition of superficial patterns etched in the
substrate is used to locate the structures. Once a satisfactory AFM image is recorded spectroscopy
measurements are performed in STM mode at several positions along the mapped structure.

AFM pictures of four of the SNS structures that were measured are shown in Fig.~2. They consist of
a loop formed with a nominally 60 nm-thick, 200 nm-wide U-shaped superconducting aluminum (Al)
electrode and a nominally 50~nm-wide and 30~nm-thick normal silver (Ag) wire. The Al film is
designed to overlap the Ag wire at both ends over nominally 50~nm. Several structures were
fabricated simultaneously on the same insulating substrate chip with Ag wire lengths ranging from
300 nm to 3~$\mu $m (for details on the fabrication see \cite{TheseHelene,AdditionalMaterial}). The
aspect ratio of the structures was chosen such that a magnetic flux $\Phi $ threading the loop
imposes a phase difference $\varphi \simeq 2\pi \Phi /\Phi _{0}$ across the silver wire
\cite{aspectratio}, with $\Phi _{0}=h/2e$ the flux quantum. To sink the tunnel current when
performing spectroscopy, the Al electrode is connected to ground on one side. The visible part of
the Ag wires is covered by a fully oxidized 2 nm-thick aluminum film. By putting the tip in
mechanical contact with this oxide layer, one can obtain nearly drift-free tunnel contacts with
conductance of the order of 1~$\mu$S. This allows to take reproducible, high resolution spectra, by
measuring the differential conductance as a function of the DC tip-sample bias voltage, using a
lock-in amplifier with typically a 1~$\mu$V AC excitation.

When the tip is placed on top of an Al electrode far away from any contact
to the Ag, the differential conductance (Fig.~2a curve A, and Fig.~2b curve
F) displays the well known behavior of a Bardeen--Cooper--Schrieffer (BCS)
superconductor, with an energy gap and its characteristic singularity at the
gap edge. These curves are well fitted with a BCS DoS, with an energy gap $%
\Delta =170~\mathrm{\mu }eV$ for Al, and an effective electronic temperature $%
T_{e}=65\,\mathrm{mK}$. This is higher than the temperature of the
refrigerator during the measurements (35~mK), probably because of a still
insufficient reduction of the electrical noise. The corresponding energy
resolution of $20~\mu \mathrm{eV}$ is nevertheless, to our knowledge, the
best obtained so far in STM spectroscopy.

\begin{figure}[tbph]
\begin{center}
\includegraphics[width=1\columnwidth,clip]{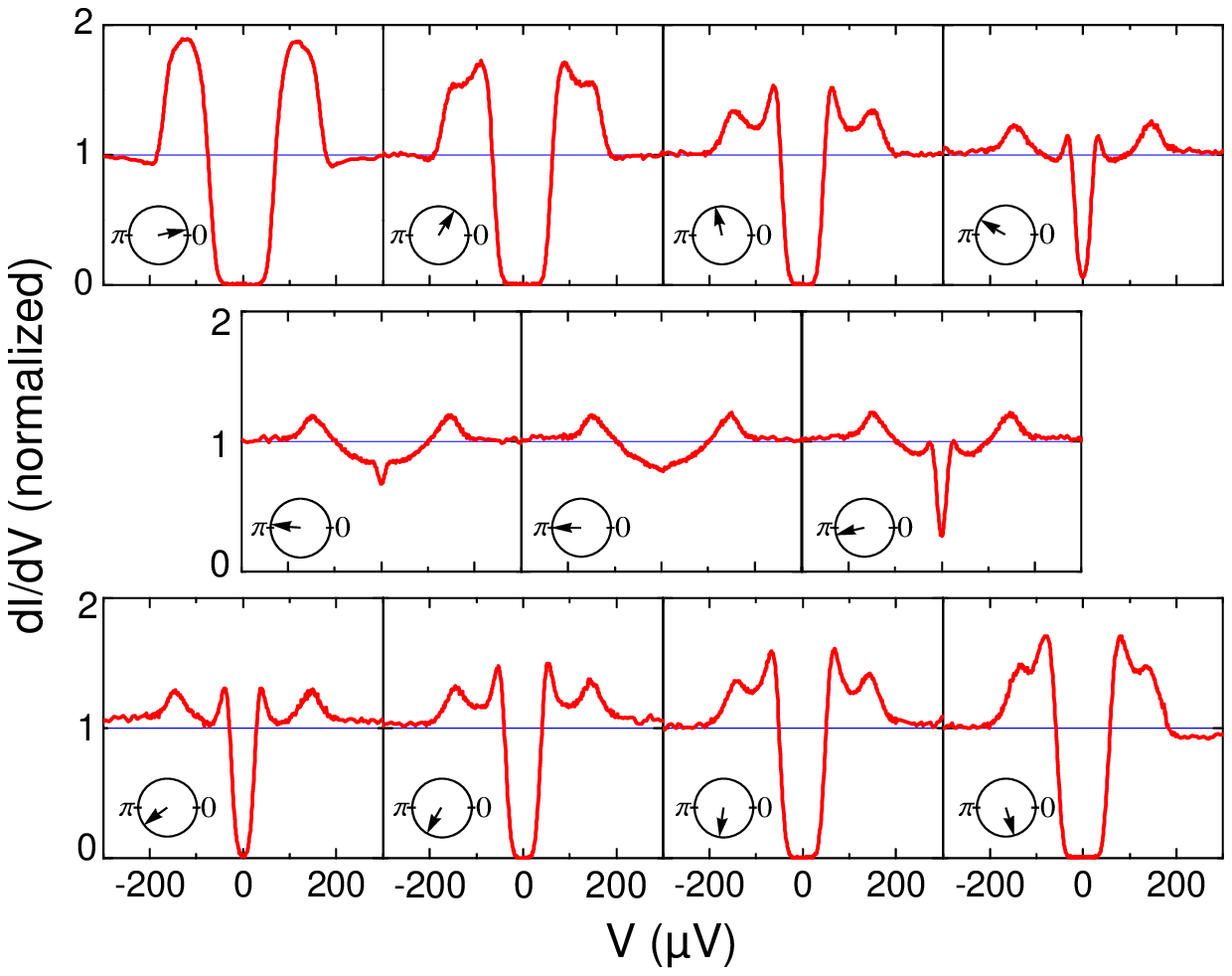}
\end{center}
\caption{Differential conductance $dI/dV(V)$ versus voltage measured in the middle of the 300
nm-long Ag wire (position D) for different values of the
phase difference across it as shown by the phase-clocks ($\protect\varphi /%
\protect\pi =0.06$, 0.32, 0.57, 0.83; 0.96, 1.00, 1.08; 1.21, 1.34, 1.46, 1.59).}
\end{figure}

\begin{figure}[tbph]
\begin{center}
\includegraphics[width=1\columnwidth]{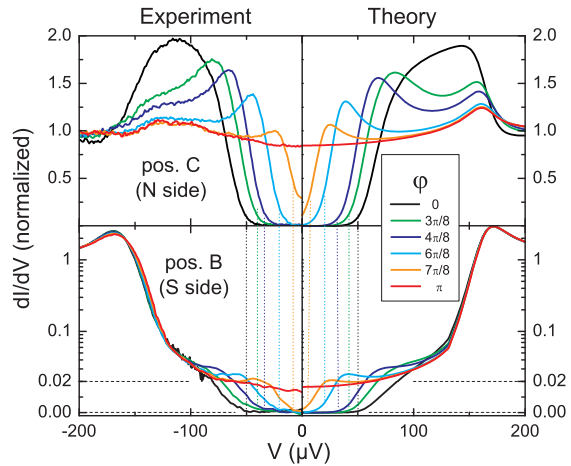}
\end{center}
\caption{(color online) Differential conductance versus voltage, for different values of the phase
difference $\protect\varphi $ across the 300 nm-long Ag wire. Left quadrants : measured; Right
quadrants : calculated. Top : on the N side (position C); Bottom : on the S side, close to the
interface (position B - Note that the scale is linear up to 0.02 and logarithmic above, to magnify
the variation for small subgap conductance). This shows that the minigap is also present on the
superconductor side with the same value as on the N side. Dashed lines are guides to the eye.}
\end{figure}

In Fig.~2 we also show the LDoS measured at zero flux for several positions
along each of the four normal wires. Both in the 300 and in the 500 nm-long
Ag wire, the differential conductance vanishes around zero voltage, on an
energy range independent of position, which is the signature of a minigap in
the DoS. The minigap is larger for the shorter wire. In the 900 nm-long
wire, the conductance is strongly reduced on a smaller energy range around
zero, but remains always finite. Finally, in the middle of the $2.7~\mathrm{%
\mu m}$-long wire, $dI/dV(V)$ is nearly constant, \textit{i.e.} unaffected
by the proximity effect. Yet, in this long wire, the LDoS is strongly
affected near the ends of the wire. When the tip is placed on the Al near a
contact to the Ag wire, a small finite conductance is observed at subgap
voltages (see \textit{e.g.} curve G in Fig.~2b, curve T in Fig.~2c and curve
B in Fig.~2a and the corresponding close-up in the bottom part of Fig.~4).
In structures displaying a finite minigap in the N wire we find that the
same minigap is also present on the S side, despite very different overall
shapes of the LDoS. Correlatively, the order parameter is slightly reduced
close to the interface as shown by a shift in the position of the peak in
the density of states ($160~\mu \mathrm{V}$ at position B \textit{vs.} $%
170~\mu \mathrm{V}$ at position A in Fig.~2a). Farther from the normal metal
(positions A and F in Fig.\thinspace 2), the subgap LDoS vanishes.

The variations with the phase difference $\varphi $ of the LDoS in the middle of the shortest wire
(position D) are shown in Fig.~3. As $\varphi $
is increased, $dI/dV(V)$ gradually deforms and $\delta$ disminishes. Around $%
\varphi =\pi $, the minigap vanishes, and the LDoS approaches that of the
normal state: superconductivity is then maximally frustrated in the wire.
This behavior is $\Phi _{0}$-periodic with the applied flux \cite%
{AdditionalMaterial}, which provides the phase calibration for each
structure. The left quadrants of Fig.~4 show the measured variations of the
subgap conductance on both the S and N side (positions B and C
respectively), establishing further that for all values of $\varphi $, the
minigap is the same in both the S and N part of the structure. A similar
phase modulation is observed in the 500 and 900~nm-long wires.

The above measurements can be compared with the predictions of the quasiclassical theory of
superconductivity in diffusive metals. This theory is based on the evaluation of the quasiclassical
retarded Green function which, for a superconductor, is a $2\times 2$ matrix of disorder-averaged
correlation functions in electron-hole (or Nambu) space. This matrix depends on position and energy
and obeys a diffusion equation, known as the Usadel equation, with boundary conditions at the
interfaces\thinspace \cite{Nazarov}. Finding the LDoS in our structures requires, in principle, to
specify their full 3-D geometry (including the interfaces and their properties), a complex task
which, as far as we know, has never been done. For simplicity, we restrict here to solving a crude
one-dimensional SNS model involving geometrical and material parameters common to all structures
(as they are fabricated simultaneously), and per-structure parameters characterizing the NS
interfaces~\cite{AdditionalMaterial}. Due to the finite extent of the overlap interfaces of N and S
in the samples, one cannot define exactly the tip-to-interface distance. Hence, for simplicity, in
the model the effective position of the tip along the effective N wire length is crudely mapped to
the apparent position of the tip on the AFM image, with an offset roughly equal to half the overlap
length (see~\cite{AdditionalMaterial} for further details). Finally, in order to account for the
effective temperature of the tip, we compare the measurements with the outcome of this model
convolved with the derivative of Fermi function at 65~mK. In spite of the simplifying assumptions
of the model, an overall agreement is obtained for the phase, space, wire length and energy
dependence of all the conductance curves, as shown in Figures 2 and 4. Deviations are most
pronounced in the shortest wire at positions near the wire ends (\textit{e.g.} positions C and E),
for which the 1D approximation neglecting the finite extent of the overlap junction geometry is
relatively more important. Also, a narrow dip in the conductance predicted near $V=0$ at the middle
of the 2.7~$\mu $m-long wire is not observed (see Fig.~2d), which may point to some uncontrolled
depairing mechanism not taken into account in the model.

In conclusion, we have measured with unprecedented energy resolution the spatial and phase
dependence of the superconducting proximity effect in a series of diffusive structures. In
particular, we have shown that an identical phase-dependent minigap is present in both the N and S
electrodes. This works complements our understanding of the proximity effect and supports its
description by the quasiclassical theory. These findings are a first illustration of what can be
achieved with this low temperature AFM-STM. There is a large variety of hybrid nanocircuits, which
combine for example magnetic materials, semiconductors or even molecules such as carbon nanotubes,
that should also give rise to a number of interesting short range ``proximity effects'' at
interfaces. In addition, the capabilities of this new instrument can be extended in various ways.
For instance, a
superconducting tip would allow probing the superconducting condensate \cite%
{Proslier} or the electronic distribution functions in nanocircuits driven
out-of-equilibrium, providing insight on local electron dynamics in the
samples \cite{Pothier}. One could also use spin-polarized tips \cite{Bode}
to resolve spin order, or use the sensor for local gating or electrostatic
force measurements \cite{Woodside} on the circuit. This combined AFM-STM
cryogenic microscope is thus a versatile tool that opens broad perspectives
in mesoscopic physics.

\begin{acknowledgments}
We gratefully acknowledge technical support by P.-F. Orfila and P. Senat and
help from the whole Quantronics group. We are also indebted to N. Agra\"{\i}%
t, H. Bouchiat, C. Chapelier, H. Courtois, S. Gu\'{e}ron, T. Heikkil\"{a},
G. Rubio-Bollinger, E. Scheer, H. Suderow and Y. de Wilde for useful
exchanges. We particularly thank J.C. Cuevas for guidance on theory and
numerics. We acknowledge support by ANR-07-Blan-0240-01 and AC Nano 2003
NR110.
\end{acknowledgments}


\begin{thebibliography}{99}
\bibitem{aspectratio} Within correcting factors of the order $\left( L_{S}+L_{G}\right) /L_{N},$ where $L_{N}$
and $L_{S}$ are the kinetic inductance of respectively the normal and the superconducting wires and
$ L_{G}$ is the geometrical inductance of the complete loop. For our samples these corrections are
always smaller than 10\%. See \cite{AdditionalMaterial, TheseHelene} for details.
\end{thebibliography}
\end{document}